\documentstyle[sprocl,psfig]{article}

\def\lsim{\mathrel{\rlap{\lower4pt\hbox{\hskip1pt$\sim$}}
    \raise1pt\hbox{$<$}}}         
\def\gsim{\mathrel{\rlap{\lower4pt\hbox{\hskip1pt$\sim$}}
    \raise1pt\hbox{$>$}}}         

\def\be{\begin{equation}}
\def\ee{\end{equation}}
\def\bq{\begin{eqnarray}}
\def\eq{\end{eqnarray}}

\mathchardef\alpha="710B
\mathchardef\beta="710C
\mathchardef\gamma="710D
\mathchardef\delta="710E
\mathchardef\epsilon="710F
\mathchardef\zeta="7110
\mathchardef\eta="7111
\mathchardef\theta="7112
\mathchardef\iota="7113
\mathchardef\kappa="7114
\mathchardef\lambda="7115
\mathchardef\mu="7116
\mathchardef\nu="7117
\mathchardef\xi="7118
\mathchardef\pi="7119
\mathchardef\rho="711A
\mathchardef\sigma="711B
\mathchardef\tau="711C
\mathchardef\upsilon="711D
\mathchardef\phi="711E
\mathchardef\chi="711F
\mathchardef\psi="7120
\mathchardef\omega="7121
\mathchardef\varepsilon="7122
\mathchardef\vartheta="7123
\mathchardef\varpi="7124
\mathchardef\varrho="7125
\mathchardef\varsigma="7126
\mathchardef\varphi="7127
\mathchardef\nabla="7272
 
\font\dozeb=cmmib10 scaled \magstep1
\font\dozesyb=cmbsy10 scaled \magstep1
\font\dezb=cmmib10
\begin{document}

\title{ A NEW GLOBAL ANALYSIS OF DIS DATA\\
AND THE  STRANGE SEA DISTRIBUTION\footnote{To be published
in the Proceedings of the Workshop on Light-Cone QCD 
and Nonperturbative Hadron Physics (Adelaide, 1999), 
World Scientific.}\\}

\author{V.~Barone}

\address{Universit{\`a} ``A.~Avogadro'', 15100 Alessandria, Italy \\
and 
INFN, Sezione di Torino, 10125 
Torino, Italy} 

\author{C.~Pascaud and F.~Zomer}

\address{LAL, IN2P3-CNRS 
 and Universit\'e de Paris-Sud,
 91898  Orsay, France}

\maketitle\abstracts{ We present the results of a new 
global analysis of DIS data, characterized by an enlarged 
neutrino and antineutrino data set. Special emphasis is 
given  to the strange sector. The strange sea distribution 
is determined independently of the non-strange sea. 
The possibility of a charge asymmetry, $s(x) \neq \bar s(x)$, 
is  tested.}

\section{Introduction}

Neutrino DIS measurements play a relatively minor role  
 in most of the present-day global fits.    
Since 
the determination of the strange sea density 
  relies completely on  charged-current DIS, $s(x)$ 
turns out to be quite poorly known. 
Recently we tried to improve the situation by 
performing a new analysis of DIS data
with an enlarged neutrino and antineutrino data set~\cite{nous}. 
This enabled us to provide an accurate 
determination of $s(x)$ in the framework of a global fit. 
We also tested the hypothesis of the 
charge asymmetry of the strange sea. 
In the following we shall briefly outline 
the results of our analysis focusing in particular 
on the strange sector.

\section{Neutrino measurements and global fits}

To start with, 
let us  see why  so far only a 
 small part of the information accumulated in 
$\nu,\bar \nu$ DIS experiments has been exploited in 
global fits.  

The old data (BEBC, CDHS, CDHSW)
cannot be immediately used because the radiative corrections
are either incomplete or not applied at all and/or the bin center
corrections are not performed. 
On the other hand, more recent data (CCFR) 
 are, so to speak, analyzed `too much'. The cross sections
are not available and only structure functions are provided. 
The latter 
come from a preanalysis, which includes assumptions on 
$R = \sigma_L/ \sigma_T$, nuclear effects, slow rescaling for charm etc. 
This can lead to  problems of consistency with the 
QCD analysis eventually performed within the global fits.

The most sophisticated global parametrizations on the market 
(CTEQ~\cite{cteq}, MRST~\cite{mrst}) use only the $F_2^{\nu}$ and 
$x F_3^{\nu}$ structure functions from CCFR. Two difficulties 
arise. First of all, the CCFR structure functions are not 
compatible at low $x$ with the charged-lepton structure functions. 
In particular, below $x \simeq 0.1$, there is a clear discrepancy 
between the CCFR $F_2^{\nu}$ and the $F_2^{\mu}$ measured 
by NMC. 
This discrepancy might be due to a different nuclear shadowing 
in $\nu$DIS and $\mu$DIS and to a different longitudinal-to-transverse
cross section ratio in charged-current and neutral-current 
DIS. 
The second problem arising from the use of CCFR structure functions
(especially, of $F_2^{\nu}$) is that they tend to favor an 
unrealistically large value of $\alpha_s$. The MRST global 
fit clearly shows that the function $\chi^2(\alpha_s)$ 
has no minimum below $\alpha_s(M_Z^2) = 0.123$. 

The difficulty of including  neutrino data 
in global fits reflects itself  in an uncertain  
knowledge of $s(x)$. 

\section{The determination of the strange sea 
density}

There are two ways to extract the strange sea distribution
from DIS data: 
{\it i)}~by a direct determination; {\it ii)}~by a global 
fit (like the other flavor densities). 

Direct determination means that the strange-charm sector 
of charged-current DIS is selected either by looking 
at particolar signatures in the final state or by 
properly combining the inclusive structure functions. 

The opposite-sign dimuon production in $\nu, \bar \nu$ DIS 
probes the strangeness in the proton.  In principle this is 
a very effective way to determine $s(x)$. However 
the analysis of this process is affected by many uncertainties
(charm fragmentation, acceptance correction, etc.). 
The dimuon determination has been performed by 
CCFR~\cite{ccfr-strange}.  
Their data sample consists of about 5,000 $\nu$ events 
and 1,000 $\bar \nu$ events. The strange-to-non-strange 
momentum ratio $\kappa$ found in the NLO CCFR analysis 
is $\kappa =0.48$ at $Q^2 = 20$ GeV$^2$.

The determination of $s(x)$ within a global fit
is made difficult by the lack of large and reliable 
neutrino and antineutrino data sets. 
Both MRST and CTEQ are unable to fit $s(x)$ independently 
of the non-strange sea. Thus they borrow the value 
$\kappa \simeq 0.5$ from the CCFR analysis and 
constrain $s(x)$ as $ s(x) + \bar s(x) = 0.5 [ \bar u(x) 
+ \bar d(x) ] $. 
The results of the fits for the 
strange distribution are clearly biased by this 
constraint and the $s(x)$  found by MRST and CTEQ depends ultimately  
on the CCFR dimuon measurement.

\section{A new global analysis and the extraction of 
$s(x)$}

We performed 
a new global analysis of DIS data at the 
NLO level in QCD. The main features of this analysis are
(for details see Ref.~1): 
\begin{itemize}
\item

The analysis is based on a large $\nu, \bar \nu$ data set, which includes 
all available 
$\nu$, $\bar \nu$ cross section data (BEBC, CDHS, CDHSW).  
We also fit charged-lepton data (BCDMS, NMC, H1) and  Drell-Yan data 
(E605, NA51, E866).
For the sake of consistency, the CCFR structure function 
data, coming from a preanalysis different from ours, are 
not included. 

\item
The data have been properly reanalyzed. Bin center corrections, 
electroweak radiative corrections, corrections for nuclear 
and isoscalarity effects have been applied. 

\item
Error correlations have been taken into account. 

\item
A massive factorization scheme is used: we chose the 
Fixed Flavor Number Scheme, which is known to be 
more stable in the moderate $Q^2$ range of 
the data.  

\item
The kinematic cuts imposed are: $Q^2 \ge 3.5$ GeV$^2$, 
$W^2 \ge 10$ GeV$^2$ (in this region higher-twist 
effects are negligible). As for the CDHSW data, we excluded 
the controversial region $x < 0.1$. 

\end{itemize}

Due to the abundance of neutrino data, we are able to 
fit the strange sea {\em independently} from the non-strange
sea. Moreover, 
the good balance between $\nu$ and $\bar \nu$ measurements  
in the high-statistics 
CDHSW data allows us to test the charge asymmetry in the 
strange sea: $s(x)  \neq \bar s(x)$ (see next section).

In our main fit, called $\tt fit1$, the parton densities are 
parametrized at the input scale $Q_0^2 = 4$ GeV$^2$ imposing 
as usual $s = \bar s$, but without any strong constraint between $s, \bar s$ 
and $\bar u, \bar d$. 
We will present the results for the strange sector, referring the reader 
to Ref.~1 for a full account of the fit.

\begin{figure}[ht]
\hspace{1.7cm}
\psfig{figure=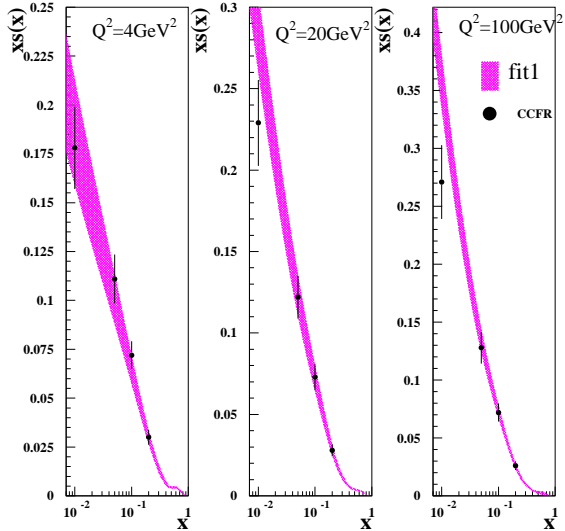,height=3.2in}
\caption{The strange distribution in {\tt fit1} at various $Q^2$ values.}
\end{figure}

In Fig.~1 we show the strange  distribution at 
different $Q^2$ values. The results of the 
CCFR dimuon determination~\cite{ccfr-strange}
is shown for comparison. 
The meaning of the error 
bands is explained in detail in Ref.~1. They 
correspond to an increase of the $\chi^2$ by one unit and 
do not take into account the uncertainties related to the 
choice of the functional form of the distributions.  
Note that although the CCFR points 
seem to be in good agreement with our curves, the 
strange-to-non-strange ratio we find is quite 
different from CCFR's: $\kappa = 0.67$ at $Q^2 = 20$ GeV$^2$, 
to be compared with the CCFR value 0.48 at the same scale. 
We also performed a modified fit, called {\tt fit1b}, 
 imposing, as it is done by CTEQ and MRST, 
 the condition $s + \bar s = 0.5 (\bar u + \bar d)$, 
 motivated by the CCFR result on $\kappa$. 
It turns out that {\tt fit1b} is definitely 
worse (see Tab.~1). 
We found that it is especially
the $\bar \nu$ 
data which favor {\tt fit1} with respect to {\tt fit1b}.

\begin{table}[hb]
\caption{The $\chi^2$'s of the three fits described in the text.}
  \begin{center}
\vspace{0.2cm}
    \begin{tabular}{|c|c|c|c|}
\hline
$\#$ pts &  $\chi^2$ {\tt fit1}  & $\chi^2$ {\tt fit1b}  
&$\chi^2$ {\tt fit2}  \\\hline
 2657 & 2430.8& 2492.4 & 2405.0\\ \hline
    \end{tabular}
  \end{center}
\end{table}

Incidentally, we notice that no discrepancy whatsoever  
emerges in our fit between neutrino and charged-lepton data. 
We  checked that the fit worsens if the CCFR structure 
functions data are taken into account.

\section{Test of the charge asymmetry in the strange sea}

Is $s(x)$ different in shape from $\bar s(x)$? 
In order to answer this question
 by fitting  neutrino and antineutrino data one needs  
 a good balance between 
$\nu$ and $\bar \nu$ statistics. 
This is the case of our data set. 

\begin{figure}[ht]
\hspace{1.7cm}
\psfig{figure=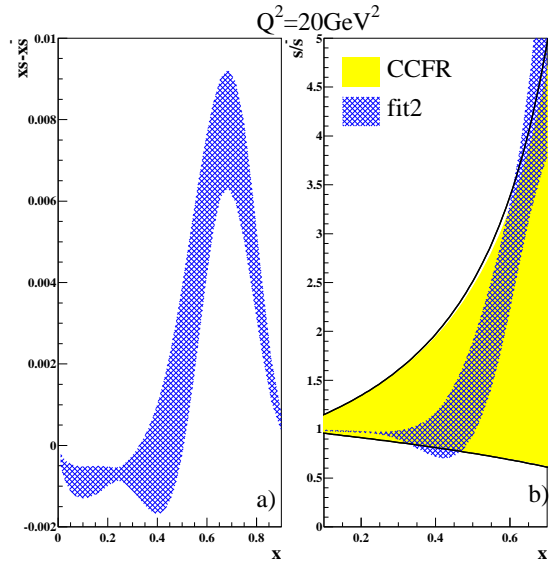,height=3.2in}
\caption{The strange vs. the antistrange distribution in {\tt fit2}.}
\end{figure}

A charge asymmetry in the strange sea is not forbidden by 
first principles (clearly, as the nucleon has no net strangeness, 
one must have $\int {\rm d}x \, (s - \bar s) =1$), and is 
actually expected in the framework of the intrinsic sea 
theory of Brodsky et al.~\cite{brodsky}. 
Intrinsic $q \bar q$ pairs have a relatively long lifetime 
and arrange themselves into higher Fock states
of the proton $\vert uud q \bar q \ldots \rangle$. By minimizing 
the kinetic energy on the light-cone one finds that the larger
the  mass of the intrinsic quark 
the higher its average momentum. Thus the
intrinsic sea tends to occupy the large $x$ region. In the specific 
case of the strange sea, the $s \bar s$ pairs give rise to 
$N \to \Lambda K$ fluctuations \cite{Lambda}. 
In Ref.~7
it was shown, by simple chiral symmetry arguments,  
 that one should expect
$\langle x_s \rangle > \langle x_{\bar s} \rangle$. 

In order to test the charge asymmetry of the strange sea, 
we released the constraint $s = \bar s$ and performed 
another fit, {\tt fit2}, looking for a possible 
difference between $s(x)$  
and $\bar s(x)$. In Fig.~2 we plot $xs(x) - x\bar s(x)$
and $s(x)/\bar s(x)$ at $Q^2 = 20$ GeV$^2$. The strange 
distribution turns out to be harder than the anti-strange one, 
in agreement with the expectation of the intrinsic sea theory. 
In Fig.~3 we show the difference $\Delta_{\nu}$ 
between $\nu$ and $\bar \nu$ 
differential cross sections which is directly sensitive 
to $s - \bar s$: {\tt fit2} is favored at large $x$ 
with respect to {\tt fit1}.  
One can see 
in Table~1 that the 
minimum $\chi^2$ of {\tt fit2} is  25 units smaller than the   
$\chi^2$ of {\tt fit1}, with an overall number of 2657 data points. 
It is clear that new high-statistics $\nu$ and $\bar \nu$ 
data would allow to increase the significance of the result and 
to draw a more definite  conclusion.

\begin{figure}[ht]
\hspace{1.7cm}
\psfig{figure=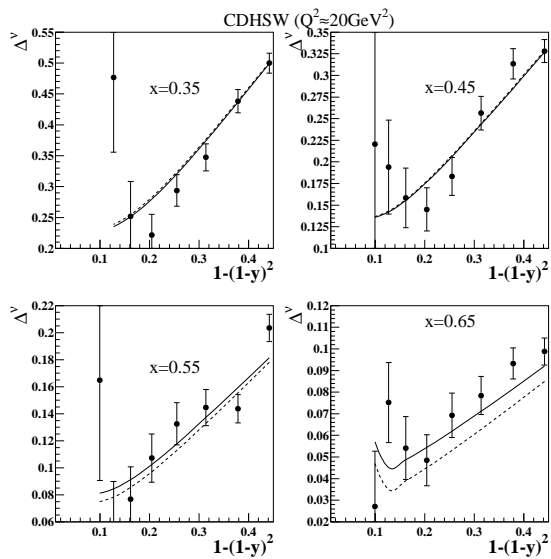,height=3.0in}
\caption{Difference between the $\nu$ and $\bar \nu$ 
differential cross sections. The solid line corresponds
to {\tt fit2}, the dashed line to {\tt fit1}.}
\end{figure}

\section{Conclusion}
 
We have shown how the full use of available $\nu, \bar \nu$ 
cross sections
provides important information on the flavor structure of the 
nucleon, in particular on the strange distribution.  
A proper analysis of the forthcoming neutrino 
 data (CHORUS and NuTeV) 
will certainly improve our knowledge of $s(x)$ and 
allow a more conclusive test of the charge asymmetry of the 
strange sea.

\vspace{1cm}


\noindent
\underline{\bf Note added.}
After we submitted this paper to the LC99 Workshop, we became aware 
of the talk
 delivered by A.~Bodek (on behalf of CCFR-NuTeV)
 at the Moriond meeting (March 2000)\cite{bodek}. 
The CCFR--NuTeV Collaboration is carrying out the analysis of new 
$\nu, \bar \nu$ DIS measurements at 
Fermilab. Their preliminary results show that:

\begin{itemize}

\item
 it is crucial not to introduce any theoretical 
bias in extracting neutrino structure functions 
from cross sections (as we stressed in Ref.~1 and in the 
present contribution); 

\item
a NLO analysis based 
on a massive scheme is needed  
in order to get reliable results, especially
in the charm-strange sector 
(as it was anticipated longtime ago\cite{tung,noi}), 
 and therefore 
  the slow rescaling 
 procedure should be definitely discarded;

\item 
 a large discrepancy emerges between the 
newly determined $F_2$ and the old CCFR $F_2$: 
this is a further,  {\it a 
posteriori}, justification for our choice of 
not using the old CCFR 
data on $F_2$ in our fits; 

\item
 the disagreement between 
$\mu$ and $\nu$ data is solved by a consistent 
 treatment of the data 
(which confirms what we claimed in Ref.~1);

\item
 measuring  the difference $xF_3^{\nu} - x F_3^{\bar \nu}$
is a viable method to extract the strange sea distribution
(the advantages of this method were pointed out, from a theoretical
 viewpoint, in Ref.~11).   

\end{itemize}

The hopefully imminent release of the CCFR-NuTeV cross section 
data will allow to push the program of Ref.~1  
 forward in the direction of a better understanding 
of the flavor structure of the proton. 


\end{document}